\documentclass[prl,twocolumn,floatfix,preprintnumbers,amsmath,amssymb,superscriptaddress]{revtex4}

\usepackage{graphicx}
\usepackage{amsmath}
\usepackage{amssymb}
\usepackage{color}
\usepackage{dcolumn}
\usepackage{epsfig}
\usepackage{bm}
\usepackage{subfigure}
\usepackage{booktabs}
\usepackage[urlcolor=blue]{hyperref}
\usepackage{makecell}
\usepackage{booktabs}
\hypersetup{colorlinks=true, linkcolor=blue, citecolor=blue}
\usepackage{multirow}

\begin{document}
	
\title{Robust superconductivity near constant temperature in rubidium-doped C$_{60}$}

\author{Li-Na Zong}
\thanks{These authors contributed equally to this work}
\affiliation{Center for High Pressure Science and Technology Advanced Research, Shanghai 201203, China}

\author{Ren-Shu Wang}
\thanks{These authors contributed equally to this work}
\affiliation{School of Science, Harbin Institute of Technology, Shenzhen 518055, China}
\affiliation{Center for High Pressure Science and Technology Advanced Research, Shanghai 201203, China}
	
\author{Di Peng}
\affiliation{Center for High Pressure Science and Technology Advanced Research, Shanghai 201203, China}
		
\author{Xiao-Jia Chen}
\email{xjchen2@gmail.com}
\affiliation{Center for High Pressure Science and Technology Advanced Research, Shanghai 201203, China}
\affiliation{School of Science, Harbin Institute of Technology, Shenzhen 518055, China}
	
\date{\today}

\begin{abstract}
To establish the doping-dependent phase diagram in alkali-metal doped C$_{60}$, we synthesize Rb-doped C$_{60}$ samples with different stoichiometries by using the improved wet-chemistry technique. The doping levels determined from the Raman scattering spectra often show the appearance of three electrons corresponding to the band filling of three for the synthesized compounds no matter matter what dopants are used. The multiple phase coexistence with the unique Rb$_{3}$C$_{60}$ is identified from the refined x-ray diffraction patterns. The phase fraction of Rb$_{3}$C$_{60}$ is found to behave with the doping in a similar manor as the superconducting shielding fraction. These rigorously established correlations among the superconducting transition temperature along with the structural and phonon vibrational properties allow us to single out Rb$_{3}$C$_{60}$ as the only superconducting phase with the nearly constant transition temperature regardless the doping level. These findings provide an experimental constraint on the theory developments for the superconductivity in fullerides.
\end{abstract}

\maketitle

The discovery of superconductivity in alkali-metal doped C$_{60}$ has stimulated great interest in exploring the chemical and physical properties of these amazing molecular superconductors. Within one year, the critical temperature ($T_c$) of fulleride superconductors has been raised from 18 to 33 K \cite{Heb,Ross,Tani}; till now, the highest record of $T_c$ at about 40 K has been achieved in Cs-doped C$_{60}$ under pressure \cite{Pals1,Ganin1,taka,Ganin2}. In the early stage, it has been discovered that $T_c$ is in a linear relation with the lattice parameter of alkali-metal doped fullerides \cite{Flem1,Zhou1,Yild1}, ascribing the family to conventional BCS superconductors, in which $T_c$ is primarily tuned by the density of states at Fermi level. Chemical and physical pressure are two prevalent methods of tuning such a parameter in alkali fullerides, and numerous works have been reported in this area \cite{Yild1,Chen,Tani2,Sparn}. Besides, tuning carrier concentration is another method to affect the state of states but has failed in the exploration for K-doped and Rb-doped binary C$_{60}$ \cite{Chen,Holc}. Interestingly, such a correlation between $T_c$ and carrier concentration was established in two cubic fullerides, Na$_2$Cs$_x$C$_{60}$ and M$_{3-y}$Ba$_y$C$_{60}$ (M = K, Rb, or Cs) \cite{Yild2}. A peaked $T_c$ was observed at or very near the half-filling of $t_{1u}$ band ($n$ = 3) in these compounds with the decreasing tendency on either side, resembling the evolution of $T_c$ with hole doping in cuprates \cite{Keimer}, which cannot be explained within the conventional BCS theory but can be well reproduced if the electron-electron and electron-phonon interactions are treated on an equal footing by using the dynamical mean-field theory \cite{Han}. The cooperation of the electron-phonon coupling and the electron correlations has been increasingly focused and explored since the discovery of the dome-like evolution of $T_c$ with pressure in Cs-doped fullerides \cite{Capone,Ganin1,Ganin2,Zadik}.

Experimentally, the controversies remain regarding whether A$_3$C$_{60}$ (A=K, Rb, Cs) is the single superconducting phase in alkali-metal doped fullerides \cite{Steph1,Murph,Fisch} or $T_{c}$ changes with the doping level $x$ in a parabolic way in A$_x$C$_{60}$. For K (Rb)-doped C$_{60}$, it has been found that the $T_c$ does not change with increasing alkali metal doping composition $x$ \cite{Chen,Holc}. For Li$_x$CsC$_{60}$ fullerides, $T_c$ was reported to change with variable $t_{1u}$ band filling, which also shows a similar electron doping dependence of $T_c$ as reported in two cubic fullerides \cite{Kosaka}. For K-doped C$_{60}$ films, the effect of the carrier concentration on $T_c$ was also reported \cite{Ren}. Nevertheless, it is still doubtful whether the valence state dependence of $T_c$ is a general rule in fulleride superconductors or not. Rb-doped C$_{60}$ is the second reported superconductor in this family, which shows the highest $T_c$ of 28-29 K at ambient pressure in binary compounds \cite{Ross}. Previous works on x-ray diffraction and Raman spectroscopy measurements demonstrate that nonstoichiometric Rb intercalation can be realized in this compound \cite{Zhu1,Mitch1}. Thus, we choose Rb-doped C$_{60}$ to examine the evolution of $T_c$ with Rb doping concentrations and try to establish the correlations among $T_c$, crystal structure, and charge transfer in these compounds.

\begin{figure}[tbp]
	\centerline{\includegraphics[width=\columnwidth]{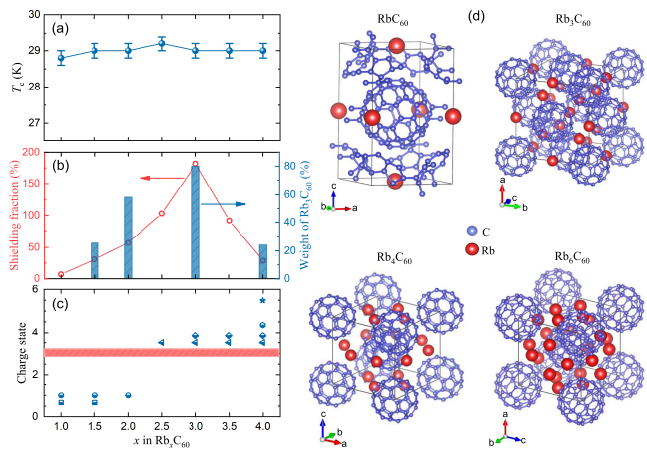}}
	\caption{(Color online) Superconducting phases in Rb$_\textit{x}$C$_{60}$ with different stoichiometries. (a) The evolution of $T_c$ with doping $x$. Error bars represent the estimated uncertainties. (b) Left axis: Plot of the superconducting shielding fraction as a function of $x$. Right axis: The phase fraction of Rb$_{3}$C$_{60}$ in the studied samples ($x$ = 1.5, 2.0, 3.0, and 4.0) obtained from x-ray diffraction analysis. (c) The evolution of charge state with $x$. The shaded region highlights the common charge state in all studied compositions. (d) Schematic crystal structure of Rb$_{x}$C$_{60}$ ($x$=1, 3, 4, and 6). The violet and red spheres represent C and Rb atoms, respectively.}
\end{figure}

The high-quality of our synthesized samples of Rb-doped C$_{60}$ is demonstrated by the successful realization of both the Meissner effect and zero-resistance state from the $dc$ [Figs. 2(a)-2(g)] and $ac$ [Fig. 2(h)] susceptibility ($\chi$) as well as resistivity [Fig. 2(i)] measurements. The imaginary part of the susceptibility ($\chi$$^{\prime\prime}$) [Fig. 2(h)] is correlated with the energy dissipation due to the formation of superconducting vertex current. A single sharp peak in $\chi$$^{\prime\prime}$ can be considered as a qualitative criterion for the realization of the zero-resistance state \cite{Gomory}, in good agreement with the resistivity data collected on the same sample as shown in Fig. 2(i). It should be mentioned that the identification of superconductivity from both the Meissner effect and zero-resistance state on the same sample rather than different ones was only completed recently for K-doped \cite{Wang,rswang2} and Rb-doped \cite{Zong} C$_{60}$ after the improvement of the synthesis technique. The early experiments for providing evidence of superconductivity either from the Meissner effect \cite{Ross,Tani} or the zero-resistance state \cite{Pals2,Mar} or rarely from both but on the different samples \cite{Heb}. So far, the essential evidence for superconductivity from the zero-resistance state for Cs$_{3}$C$_{60}$ is till missing and waits to be filled up \cite{Pals1,Ganin1,taka,Ganin2}. Holding the two essential characters of superconductivity for Rb-doped C$_{60}$ make this study confident in examining the doping effect on superconductivity compared to the early studies \cite{Yild2,Kosaka}. 

\begin{figure*}[tbp]
	\centerline{\includegraphics[width=2\columnwidth]{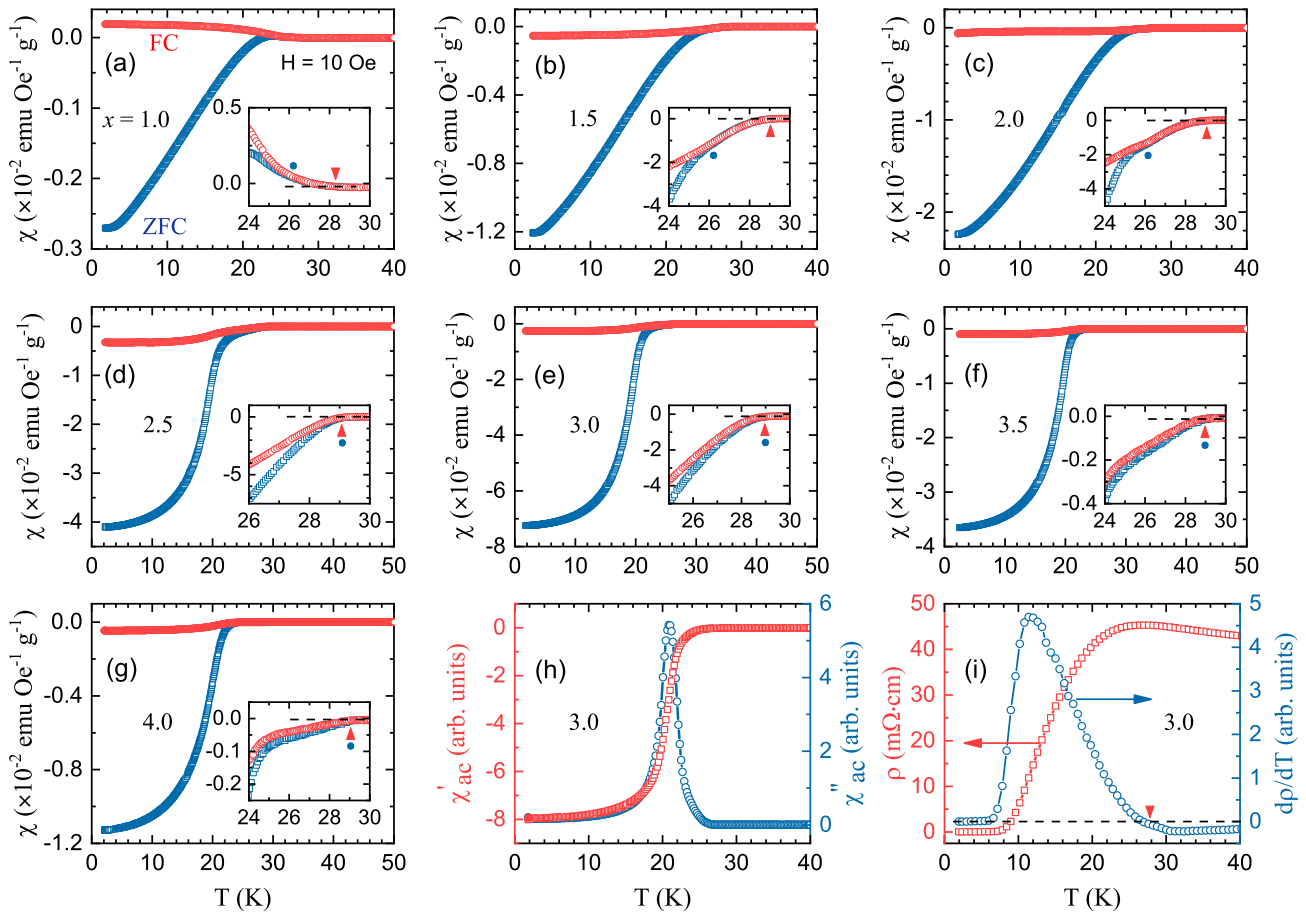}}
	\caption{(Color online) Superconductivity in nominal Rb$_{x}$C$_{60}$ ($x$=1, 1.5, 2, 2.5, 3, 3.5, and 4). (a)-(g) Temperature dependence of the $dc$ $\chi$. Inset: An expanded view in a narrow temperature range near the superconducting transition. The dashed lines represent the linear extrapolation from the normal state (30 K), and the temperatures marked by solid triangles and solid circles for each sample represent the real $T_c$ and the branching point for the ZFC and FC curves, respectively. (h) Temperature dependence of the $ac$ $\chi$ for nominal Rb$_{3}$C$_{60}$. (i) Temperature dependence of the electrical resistivity ($\rho$) and its derivative for nominal Rb$_{3}$C$_{60}$.}
\end{figure*}

Our central results for Rb$_\textit{x}$C$_{60}$ with different stoichiometries are summarized in Fig. 1. Based on the magnetization measurements, as shown in Fig. 1(a), $T_c$ resists doping concentrations within our studied stoichiometric range. The similar phenomenon was observed previously but with only a few compositions \cite{Chen}. In Fig. 1(b), the regularity obtained from the magnetization measurements for a series of samples is that the shielding fraction reaches a maximum in nominal Rb$_{3}$C$_{60}$ and decreases on either side of $x$ = 3. This phenomenon was also noticed for K-doped C$_{60}$ and Rb-doped C$_{60}$ \cite{Chen,Holc}, indicating that the superconducting phase in these compounds is close to $x$ = 3 but still lack of unambiguous evidence. According to the x-ray diffraction analysis, the right axis of Fig. 1(b) displays the phase fraction of the refined Rb$_{3}$C$_{60}$ in nominal Rb$_{1.5}$C$_{60}$, Rb$_{2}$C$_{60}$, Rb$_{3}$C$_{60}$, and Rb$_{4}$C$_{60}$. The change of the phase fraction for refined Rb$_{3}$C$_{60}$ with increasing Rb concentration is analogous to the law established for shielding fraction, namely, the phase fraction of Rb$_{3}$C$_{60}$ is peaked at nominal composition $x$ = 3 and also decreases on either side. Apart from that, based on the refinement results, we also find that, at room temperature, there are 4 stable phases in Rb$_\textit{x}$C$_{60}$ within our studied scope, including RbC$_{60}$, Rb$_{3}$C$_{60}$, Rb$_{4}$C$_{60}$, and Rb$_{6}$C$_{60}$, whose structure can be assigned to the orthorhombic, face-centered cubic ($f.c.c.$), body-centered tetragonal ($b.c.t.$), and body-centered cubic ($b.c.c.$) phase, respectively. Figure 1(d) exhibits the schematic crystal structures for the phases mentioned above, which are in accord with those reported in Ref. \cite{Flem2}. Based on the Raman scattering measurements, Fig. 1(c) depicts the overall evolution of the charge state of C$_{60}$ with increasing Rb doping concentration. Different charge states can be detected in one sample, which suggests that phase separation is a common feature for Rb$_\textit{x}$C$_{60}$. In addition, the charge state of C$_{60}$ increases gradually with increasing doping composition, in which the phase with 3 charge transfer is the common phase in all compositions. Combining the results derived from the magnetization, x-ray diffraction, and Raman spectroscopy measurements, one can safely draw a conclusion that the robust superconductivity with $T_c$ of 28-29 K resisting doping concentrations in Rb-doped C$_{60}$ is from $f.c.c.$ Rb$_{3}$C$_{60}$ with exact 3 charge transfer. Besides, phase separation is verified by both x-ray diffraction and Raman spectroscopy in all the studied Rb-doped C$_{60}$.

The shown $T_c$ and superconducting shielding fraction for each Rb$_{x}$C$_{60}$ in Figs. 1(a) and 1(b) (left axis) are based on the $dc$ magnetization measurements.  Figure 2(a)-2(g) show the temperature dependence of $dc$ $\chi$ for samples with different nominal compositions. Both the zero-field-cooling run (ZFC, blue curve) and the field-cooling run (FC, red curve) for each sample exhibit obvious diamagnetic transition, in which the ZFC and the FC runs represent the magnetic exclusion and the magnetic expulsion, respectively. The big difference between two curves is a phenomenon of flux trapping, which is a typical character of type-II superconductors. Particularly, the ZFC curve and the FC curve for nominal RbC$_{60}$ show anomalous features after the superconducting transition. The ZFC curve has a small paramagnetic bump, and the FC curve flips into weak-paramagnetic area. This phenomenon has also been observed in other work \cite{Dahl}, in which it was attributed to the systematic error caused by remanent field in the instrument, but it also may be derived from the strong flux-pinning caused by large amounts of defects or non-superconducting phases. Due to the sonication treatment in the synthesis method, polycrystalline powders are inclined to form, and the granular effect is so distinct that the superconducting transition is not narrow enough for our samples. Thus, a defining method of $T_c$ having been applied for powdered samples is adopted to give an accurate result \cite{Dahl}. The inset of each figure shows an expanded view near the superconducting transition and presents the method to determine $T_c$ for each sample. The critical temperature is defined as the first temperature point that deviates from the linear extrapolation of the normal state, as marked by a solid triangle in each inset. This temperature represents the initial point for the formation of electron pairs. Such a definition gives the consistent $T_{c}$ value for Rb$_{3}$C$_{60}$ as indicated from the $dc$ [Fig. 2(e)] and $ac$ $\chi$ [Fig. 2(h)] together with resistivity [Fig. 2(i)] measurements on the same sample.  For nominal RbC$_{60}$, Rb$_{1.5}$C$_{60}$, and Rb$_{2}$C$_{60}$ compounds, the $T_c$ is not consistent with the branching point of the ZFC and FC curves marked by the solid circle, which may result from the deteriorated superconducting phase. Assuming the density ($\delta$) of our synthesized samples is 2 g cm$^{-3}$, one can calculate the superconducting shielding fraction (SF = 4$\pi$$\chi$$\delta$) by using the magnetic susceptibility collected at 2 K with an applied field of 10 Oe. The obtained results are displayed in Fig. 1(b). For nominal Rb$_{2.5}$C$_{60}$ and Rb$_{3}$C$_{60}$, the apparent SF is larger than 100\%, which has also been discovered in previous K- and Rb-doped C$_{60}$ superconductors \cite{Irons,Buntar1}. For type-\uppercase\expandafter{\romannumeral2} superconductors, the SF is usually larger than the superconducting volume fraction, and the former is also in proportion to the latter; only in perfect superconductors, the SF is equal to the superconducting fraction and can be calculated in terms of the FC susceptibility \cite{Buntar2}. Due to the demagnetization factor, the SF could be larger than 100\%, and it is a normal phenomenon in alkali-metal doped C$_{60}$.	

\begin{figure*}[tbp]
	\centerline{\includegraphics[width=2\columnwidth]{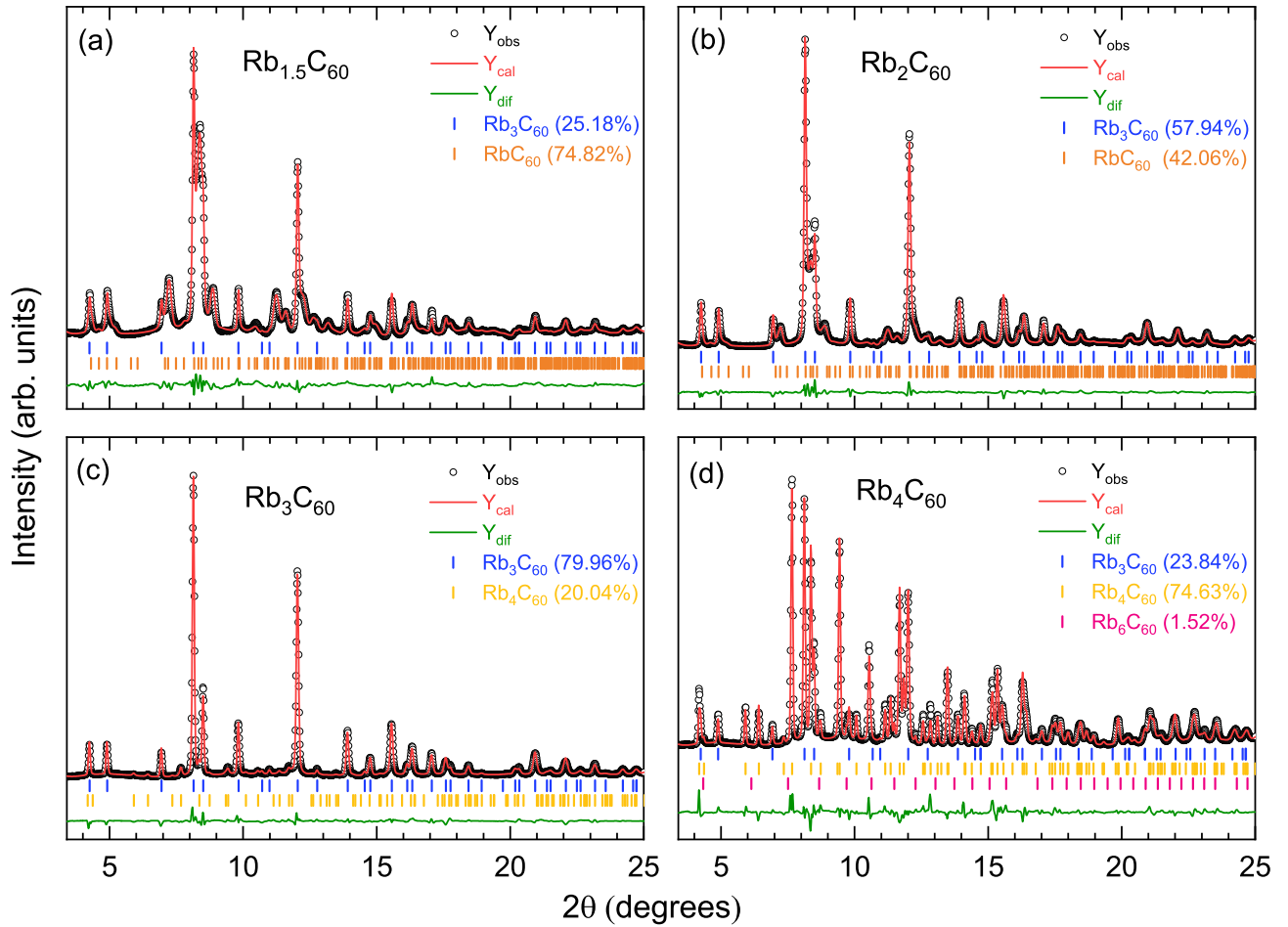}}
	\caption{(Color online) (a)-(d) X-ray diffraction spectra and Rietveld refinements of Rb$_\textit{x}$C$_{60}$ ($x$=1.5, 2, 3, and 4). Experimentally observed (open circles), calculated (red lines), and difference profiles (dark green lines) of the multi-phase refinement are given for each sample. Sticks represent the calculated reflection positions for different phases.}
\end{figure*}

The shown phase fraction of Rb$_{3}$C$_{60}$ [right axis of Fig. 1(b)] and the crystal structures [Fig. 1(d)] of the synthesized samples were determined from the well refined x-ray diffraction data for nominal Rb$_{x}$C$_{60}$ ($x$=1.5, 2, 3, and 4).  Figure 3 displays the Rietveld refinement results for each sample. The detailed refinement results including the phase fraction, space group, and lattice parameters are given in Table I. As learned, phase separation is a common feature, and all samples have at least two phases, in which nominal Rb$_{1.5}$C$_{60}$, Rb$_{2}$C$_{60}$, and Rb$_{3}$C$_{60}$ have two coexisting phases, and nominal Rb$_{4}$C$_{60}$ has three coexisting phases. With increasing Rb doping concentration, different phases with variable Rb compositions appear. In nominal Rb$_{1.5}$C$_{60}$ and Rb$_{2}$C$_{60}$, the orthorhombic RbC$_{60}$ in the space group of $Pmnn$ coexists with the $f.c.c.$ Rb$_{3}$C${60}$ with the space group of $Fm\bar{3}m$. In nominal Rb$_{3}$C$_{60}$ and Rb$_{4}$C$_{60}$, the $f.c.c.$ Rb$_{3}$C$_{60}$ coexists with the $b.c.t.$ Rb$_{4}$C$_{60}$ with the space group of $I4/mmm$, but traces of $b.c.c.$ Rb$_{6}$C$_{60}$ in the space group of $Im\bar{3}$ can also be detected in nominal Rb$_{4}$C$_{60}$. The $f.c.c.$ Rb$_{3}$C$_{60}$ is the common phase in all studied samples, whose phase fraction is peaked at nominal composition $x$ = 3, as shown in Fig. 1(b). Figure 1(d) displays the crystal structure for the four phases mentioned above. The $f.c.c.$ Rb$_{3}$C$_{60}$ is the most studied phase and the generally-accepted superconducting phase \cite{Fisch,McCau}. The crystal structure of Rb$_{4}$C$_{60}$ resembles that for Rb$_{6}$C$_{60}$ but losing two Rb atoms on the four faces parallel to the $c$ axis, and it can be regarded as a distorted $b.c.c.$ structure. 

\begin{table}[tbp]
	\caption{Rietveld refinement results of the phase and its weight $w$ (in $\%$) with the space group ($sg$), lattice parameter(s) (in \AA), unit cell volume $V$ (in \AA$^3$), weighted profile $R$-factor ($R_{wp}$ in $\%$), and goodness of fit ($Gof$ in $\%$) for nominal Rb$_{x}$C$_{60}$ ($x$=1.5, 2, 3, and 4).}
\begin{center}	
	\begin{tabular}{c|c|c|c|c|c}
		\hline
		\hline
		\makecell[c]{$x$} & \makecell[c]{$w$ ($sg$) \\} & $a$, $b$, $c$ & $V$ & $R_{wp}$ & $Gof$ \\
		\hline
		       1.5 & \makecell[c]{RbC$_{60}$-74.82\\($Pmnn$)} & \makecell[c]{$a$=9.1297(64)\\$b$=10.0495(63)\\$c$=14.483(10)} & 1328.8(16) & \multirow{2}*{7.698} & \multirow{2}*{0.729} \\
		\cline{2-4}
		& \makecell[c]{Rb$_3$C$_{60}$-25.18\\($Fm\overline{3}m$)} & $a$=14.4823(65) & 3037.0(31) &  & \\
		\hline
			2 & \makecell[c]{RbC$_{60}$-42.06\\($Pmnn$)} & \makecell[c]{$a$=9.0342(62)\\$b$=10.0841(73)\\$c$=14.480(12)} & 1319.2(17) & \multirow{2}*{5.730} & \multirow{2}*{0.718} \\
		\cline{2-4}
		& \makecell[c]{Rb$_3$C$_{60}$-57.94\\($Fm\overline{3}m$)} & $a$=14.4616(67) & 3024.3(31) &  & \\
		\hline
			3 & \makecell[c]{Rb$_3$C$_{60}$-79.96\\($Fm\overline{3}m$)} & $a$=14.4755(71) & 3032.9(33) & \multirow{2}*{8.104} & \multirow{2}*{1.150} \\
		\cline{2-4}
		& \makecell[c]{Rb$_4$C$_{60}$-20.04\\($I4/mmm$)} & \makecell[c]{$a$=12.0256(66)\\$c$=11.0380(66)} & 1596.3(20) &  & \\
		\hline
			4 & \makecell[c]{Rb$_3$C$_{60}$-23.84\\($Fm\overline{3}m$)} & $a$=14.5162(79) & 3060.8(38) & \multirow{3}*{10.776} & \multirow{3}*{1.336} \\
		\cline{2-4}
		& \makecell[c]{Rb$_4$C$_{60}$-74.63\\($I4/mmm$)} & \makecell[c]{$a$=12.0162(63)\\$c$=11.0826(58)} & 1600.2(19) &  & \\
		\cline{2-4}
		& \makecell[c]{Rb$_6$C$_{60}$-1.52\\($Im\overline{3}$)} & $a$=11.5898(63) & 1556.8(25) &  & \\
		\hline
		\hline
		\end{tabular}		
\end{center}
\end{table}

The charge state of the phase with increasing Rb concentration shown in Fig. 1(c) was determined from the analysis of the collected Raman spectra (Fig. 4). As shown in Fig. 4(a), for pristine C$_{60}$, there are six obvious Raman active modes, consisting of four radial $H_g$ modes and two tangential $A_g$ modes. Upon increasing electron doping, the downshift of the $A_g$(2) mode for one electron is 6 cm$^{-1}$, consistent with the charge transfer law \cite{Haddon1,Kuzmany}. For the phases with charge state ($n$) $\leq$ 1, there are six apparent Raman active modes like those in pristine C$_{60}$, but the relative intensity between the $H_g$(1) [$A_g$(1)] mode and $A_g$(2) mode increases apparently upon doping. Apart from that, the broadened line width for the $H_g$(7) and $H_g$(8) modes is another character for charge transfer as reported in previous works \cite{Winter1}, indicating the increased electron-phonon coupling. For the phase with exact 3 charge transfer, only 4 Raman active modes can be detected on the spectra, including the two low-energy $H_g$ modes and two high-energy $A_g$ modes. The most prominent feature for this state is the increasingly broadened $H_g$(1) and $H_g$(2) modes, which is generally recognized as a character of strong electron-phonon coupling \cite{Zhou2}. The characters detected from our spectra are consistent with those reported for Rb$_{3}$C$_{60}$ \cite{Zhou2}. It is well accepted that the phase with 3 charge transfer or half-filling $t_{1u}$ band is in a metallic state and can be attributed to the superconducting phase of alkali-fullerides \cite{Haddon2}. When the charge transfer is larger than 3, the Raman active modes are similar to those with $n$ = 3. The gradually narrowed $H_g$(1) and $H_g$(2) modes have the strong salient features, which is reminiscent of the spectra for Rb$_{4}$C$_{60}$ \cite{Mitch2}. The largest charge state of 5.5 corresponds to an early stage of the fully-doped Rb$_{6}$C$_{60}$, which has also been reported in K-doped C$_{60}$ \cite{Winter2}. In particular, the phase with $n$ = 5.5 observed at one point in nominal Rb$_{4}$C$_{60}$ is only a coexisting phase for another one with less charge transfer ($n$ = 3.5), which also indicates that only traces of Rb$_{6}$C$_{60}$ can be detected in nominal Rb$_{4}$C$_{60}$. This result is consistent with the phase fraction of Rb$_{6}$C$_{60}$ obtained from the x-ray diffraction analysis. Similar to the XRD data, phase separation is a prominent character for our Rb-doped C$_{60}$ compounds. However, the charge state of three electrons is the only feature shared for all the doped compounds. Therefore, the corresponding Rb$_{3}$C$_{60}$ is identified as the only phase to exhibit superconductivity with nearly constant $T_{c}$ upon the change of Rb concentration. This comprehensive study fails to find the desired dome shape for the $T_{c}$ change with doping in Rb-doped C$_{60}$ and hopes to provide the base on the theory development of superconductivity in fullerides. 

\begin{figure*}[tbp]
	\centerline{\includegraphics[width=2\columnwidth]{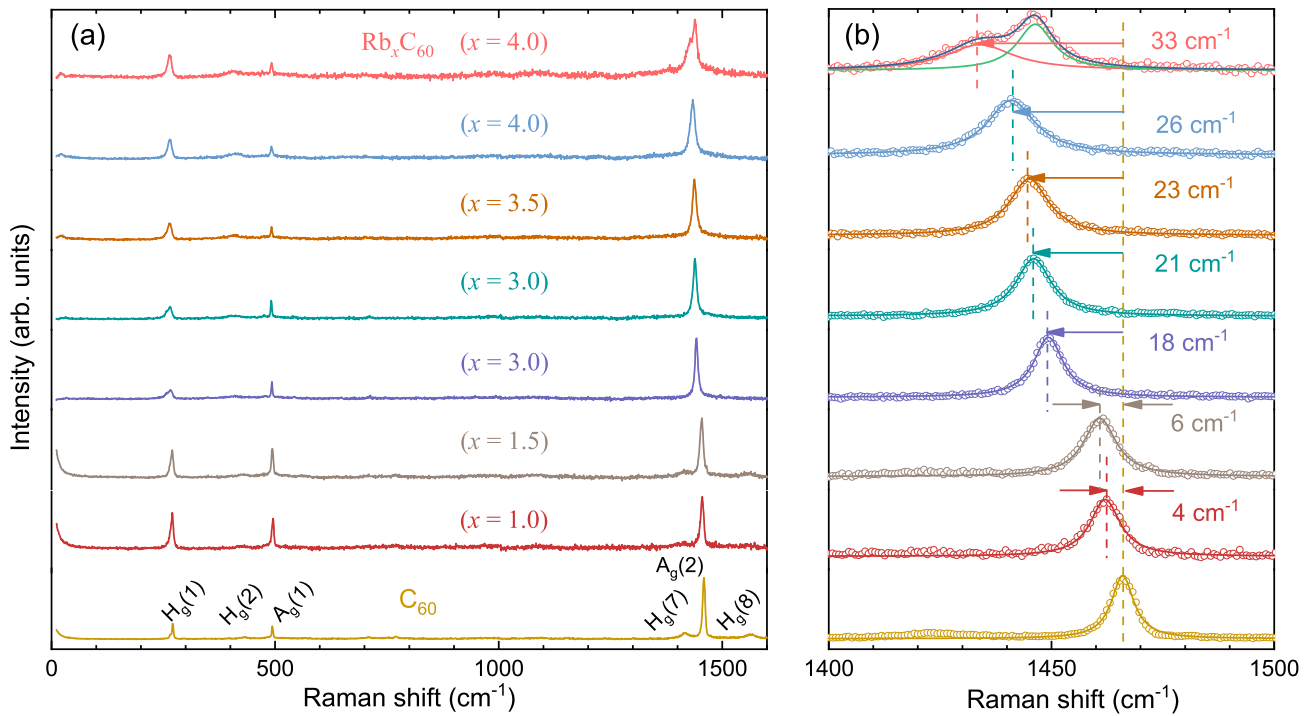}}
	\caption{(Color online) Raman spectra in the full range (a) and the expanded view of the $A_g$(2) modes (b) of pristine C$_{60}$ and Rb-doped C$_{60}$ with increasing charge transfer. The hollow circles represent the observed profiles, and the solid lines are the Lorentz fitting results.}
\end{figure*}

\vspace{0.3cm}
\noindent\textbf{Experimental details:} Pristine C$_{60}$ ($>$ 99.5\%) were purchased from Tokyo Chemical Industry (Shanghai) and used without further purification. The rubidium metal with purity no less than 99.75\% was purchased from Alfa Aesar. We adopted a similar wet-chemical method to synthesize Rb-doped C$_{60}$ samples with different compositions as that in our previous work \cite{Zong}. The pre-weighted C$_{60}$ (40 mg) and rubidium metal with different mole ratio (C$_{60}$ : Rb = 1:1.0-4.0) were first put in a borosilicate vial, and then 3-4 mL ultra-dry tetrahydrofuran solvent was poured into the vial. After that, the reaction vial sealed with a cap was transferred to an ultrasonic instrument with oil bath and was sonicated for 9-10 minutes at about 40 $^{\circ}$C. After sonication, the colorless solvent turned into dark-purple or red-brown mixtures, which depended on the mole ratio of Rb metal, and then the reaction vial was loaded on a vortex oscillator and further reacted for 6-8 h before a still standing process. After the standing process, the precipitated product was filtrated and a black precursor with solvent molecules was obtained. For samples with nominal stoichiometry in the range of 1.0-2.0, due to the high solubility of the synthesized samples, no precipitation could be obtained by the standing process. N-hexane, as a precipitation agent, was added to accelerate the precipitation and to acquire a highly crystallized sample. Annealing process was conducted at 220 $^{\circ}$C for 48 hours in an inert atmosphere followed by a natural cooling process. All the manipulations mentioned above were carried out in a glovebox filled with argon (99.999\%) and with the content of H$_{2}$O and O$_{2}$ less than 0.1 ppm.

After obtaining the final products, they were sealed in non-magnetic capsules and glass capillary tubes for magnetization, Raman spectroscopy, and Synchrotron x-ray diffraction measurements, respectively. The magnetization measurements were performed in a SQUID magnetometer (Quantum Design) with a given magnetic field of 10 Oe. Both $dc$ and $ac$ magnetic susceptibility were collected to give an overall magnetic behavior. In $dc$ magnetic susceptibility measurements, the sample was first cooled down to 2 K, and then a magnetic field of 10 Oe was applied, zero-field-cooling data were recorded while warming to 40 or 50 K. The field-cooling run was collected at the same magnetic field when the sample was cooled down to 2 K again. For the $dc$ magnetic susceptibility measurements, the probe magnetic amplitude and frequency are 1 Oe and 234 Hz, respectively. The resistivity for nominal Rb$_{3}$C$_{60}$ was also obtained by applying a $dc$ four-probe method in Van Der Pauw configuration \cite{pauw} in the Physical Properties Measurement System (Quantum Design). Raman spectroscopy was collected in an in-house system with a charged coupled device and spectrometer from Princeton Instruments. The laser with a wavelength of 488 nm and laser power less than 1 mW was adopted to avoid the possible radiation damage. The integration time for each spectrum was set to 1 min. To give an accurate result, all Raman spectra were collected under the same conditions, and we collected 5-7 spectra from different points in one sample to ensure the comparability between data. Synchrotron x-ray diffraction patterns were collected on the BL15U1 synchrotron beamline at Shanghai Synchrotron Research Facility using a focused monochromatic beam. An incident x-ray beam with wavelength of 0.6199 {\AA} and a Mar165 two-dimensional charge-coupled device detector were used in the measurement. The obtained XRD patterns were integrated and analyzed in Dioptas software package \cite{Presch}. The standard Rietveld refinement was performed to get the knowledge of the space group, the unit cell parameters, and the phase fractions for different coexisting phases using Topas software package \cite{Coel}.
	
\begin{acknowledgments}	
This work was funded by the Shenzhen Science and Technology Program (Grant No. KQTD20200820113045081), the Basic Research Program of Shenzhen (Grant No. JCYJ20200109112810241), and the National Key R$\&$D Program of China (Grant No. 2018YFA0305900).
\end{acknowledgments}

\end{document}